\documentclass[6pt,onecolumn]{article}
\usepackage[english]{babel}
\usepackage[utf8]{inputenc}

\usepackage{amsmath}
\usepackage{amsfonts}
\usepackage{amssymb}
\usepackage{graphicx}
\usepackage[affil-it]{authblk}
\usepackage{hyperref}
\usepackage{color}
\usepackage{caption}
\usepackage{verbatim}
\usepackage[normalem]{ulem}
\usepackage{subfigure}
\usepackage{slashed}
\usepackage{cancel}
\usepackage{braket}
\usepackage{multicol}
\usepackage{tcolorbox}
\tcbuselibrary{theorems}
\usepackage{bm}
\usepackage{enumitem}
\usepackage[numbers,sort&compress]{natbib}

\hypersetup{linktocpage=true}

\usepackage{tikz} 
\usepackage{tikz-feynman}

\tikzfeynmanset{compat=1.0.0}

\usetikzlibrary{arrows,shapes}
\usetikzlibrary{trees}
\usetikzlibrary{matrix,arrows} 				
\usetikzlibrary{positioning}				
\usetikzlibrary{calc,through}				
\usetikzlibrary{decorations.pathreplacing}  
\usepackage{pgffor}							

\usetikzlibrary{decorations.pathmorphing}	
\usetikzlibrary{decorations.markings}
\tikzset{
    vector/.style={decorate, decoration={snake}, draw},
	provector/.style={decorate, decoration={snake,amplitude=2.5pt}, draw},
	antivector/.style={decorate, decoration={snake,amplitude=-2.5pt}, draw},
    fermion/.style={draw=black, postaction={decorate},
        decoration={markings,mark=at position .55 with {\arrow[draw=black]{>}}}},
    fermionbar/.style={draw=black, postaction={decorate},
        decoration={markings,mark=at position .55 with {\arrow[draw=black]{<}}}},
    fermionnoarrow/.style={draw=black},
    gluon/.style={decorate, draw=black,
        decoration={coil,amplitude=4pt, segment length=5pt}},
    scalar/.style={dashed,draw=black, postaction={decorate},
        decoration={markings,mark=at position .55 with {\arrow[draw=black]{>}}}},
    scalarbar/.style={dashed,draw=black, postaction={decorate},
        decoration={markings,mark=at position .55 with {\arrow[draw=black]{<}}}},
    scalarnoarrow/.style={dashed,draw=black},
    electron/.style={draw=black, postaction={decorate},
        decoration={markings,mark=at position .55 with {\arrow[draw=black]{>}}}},
	bigvector/.style={decorate, decoration={snake,amplitude=4pt}, draw},
    line/.style={draw=black},
}\usetikzlibrary{decorations.markings}

\title{$Z_3$ Scalar Dark Matter with Strong Positron Fluxes}
\author[1]{Basti\'an D\'iaz S\'aez\thanks{bastian.diaz@tum.de}}
\author[2]{Karim Ghorbani\thanks{karim1.ghorbani@gmail.com}}
\affil[1]{Physik-Department, Technische Universität München, James-Franck-Straße, 85748 Garching, Germany}
\affil[2]{Physics Department, Faculty of Sciences, Arak University, Arak 38156-8-8349, Iran}

\begin{document}
\twocolumn
\maketitle
\begin{abstract}
We explore a class of simplified extensions to the Standard Model containing a complex singlet scalar as a
dark matter candidate accompanied by a vector-like lepton as a mediator, both charged under a new $Z_3$ symmetry. In
its simplest form, the new physics couples only to right-handed electrons, and the model is able to accommodate
the correct dark matter relic abundance around the electroweak scale up to several TeV evading the strongest
constraints from perturbativity, collider and dark matter searches. Furthermore, the model is capable to enhance
naturally positron fluxes by several orders of magnitude presenting a box-shape spectra. This framework opens up a lot of phenomenological possibilities depending on the quantum charge assignments of the new fields.
\end{abstract}


\section{Introduction}\label{int}
Over the years, simple extensions to the SM have been proposed to account for the elusive dark matter (DM) particle, with many of them presenting distinctive observational signatures for different experimental setups. One of the simplest extensions to the SM to account for DM and distinctive indirect detection signals is the class of models with two fields transforming under a discrete $Z_2$ symmetry: real scalar DM plus a vector-like lepton, or a Majorana fermion plus a scalar singlet field \cite{Bell:2011eu, Bringmann:2012vr, Bergstrom:2013jra, Garny:2013ama, Giacchino:2013bta, Toma:2013bka, Bringmann:2013oja, Giacchino:2014moa, Okada:2014zja, Agrawal:2014ufa, Kopp:2014tsa, Kelso:2014qja, Ibarra:2014qma, TOMA2015188,
 Duerr:2015vna, Bringmann:2015cpa, Giacchino:2015hvk, Kumar:2016cum, Sandick:2016zut, Baek:2016lnv, Khoze:2017ixx, Bell:2017irk, Bringmann:2017sko, ElAisati:2017ppn, Barducci:2018esg, Calibbi:2018rzv, Colucci:2018vxz, Colucci:2018qml, Biondini:2019int, Junius:2019dci, Baum:2020gjj, Arina:2020tuw, Arina:2020udz, Acuna:2021rbg, Becker:2022iso}. Focusing on the former, these class of models present distinctive gamma-ray signatures via Bremsstrahlung radiation sensitive to Fermi-LAT or even to the Cherenkov Telescope Array (CTA), but in terms of anti-particles, they lack sizable fluxes. 


Even when $Z_2$ symmetry is the simplest assumption to explore models with DM candidates, higher symmetries could shed light on interesting phenomenological aspects, without complicating things in a significant manner. For instance, a complex scalar with $Z_3$ symmetry disentangles the freeze-out relic values with direct detection observables \cite{DEramo:2010keq, Belanger:2012zr}, recovering some parameter space of singlet DM at the electroweak (EW) scale in comparison to its $Z_2$ analogue real singlet \cite{Cline:2013gha}. Ideas for fermionic DM \cite{Cai:2015zza, Guo:2021rre}, multicomponent dark matter \cite{Yaguna:2019cvp, Yaguna:2021rds, Bhattacharya:2017fid, Choi:2021yps},
complex singlet plus a second doublet scalar \cite{Belanger:2012vp}, neutrino mass generation \cite{Ma:2007gq, Qi:2021rpa}, phase transitions \cite{Chiang:2019oms, Ghosh:2022fzp}, inflation \cite{Choi:2020ara}, inert doublets \cite{Aranda:2019vda} and the positron excess \cite{Cai:2018imb} can be realized based on this new symmetry, to mention some of them.

In this work we explore a simple scenario in which we extend the SM model with two fields, a complex singlet scalar $S$ and a vector-like lepton (VLL) $\psi$, with both fields charged identically under the same $Z_3$ symmetry. Provided that $m_S<2m_\psi$, the complex singlet becomes stable, becoming a DM candidate. The simplest realization of this scenario is when the new fields are $SU(2)_L$ singlets, \textit{leptophilic} (also called \textit{lepton portal DM}) \cite{Bergstrom:2008gr, Bell:2011eu, Cavasonza:2016qem, Profumo:2019ujg, Kawamura:2020qxo}, and when the new fields couple only to right-handed electrons. We show that DM semi-annihilations play an important role in both relic density calculation and indirect detection signals. Even when direct detection appears at one-loop level, its effects are rather strong, but since the couplings in this $Z_3$ framework tend to be milder (contrary to its analogue $Z_2$ version with a complex scalar DM and a VLL mediator \cite{Chang:2014tea, Bai:2014osa, Kawamura:2020qxo}), it is possible to evade these bounds keeping DM at the GeV-TeV scale. 

Interestingly, in some regions of the parameter space, this minimal $Z_3$ leptophilic scenario may present sizable fluxes of positron fluxes featuring box-shape spectra \cite{Zu:2017dzm}, thus distinctive for astrophysical probes. We constrain the model with model-independent bounds based on AMS-02 data \cite{PhysRevLett.122.041102, Ibarra:2013zia}, showing exclusions in some portions of the parameter space below TeV, and other regions requiring more sensible experiments to be tested.

The paper is organized in the following way. In section \ref{setup} we introduce the model. In section \ref{constrains} we present the most relevant constraints on the model. In section \ref{results} we present the resulting parameter space of the model after imposing the most relevant constraints, and finally in section \ref{conc} we state the conclusions. 

\section{Model}\label{setup}
Besides the SM particle content, we consider leptophilic new physics consisting of a complex singlet scalar $S$ (along with its complex conjugate $S^*$) and a vector-like lepton (VLL) $\psi$ with $Y=-1$, both charged under a global $Z_3$: $S\rightarrow e^{i2\pi/3}S$ and $\psi\rightarrow e^{i2\pi/3}\psi$. We assume that the new sector only couples to the first right-handed lepton generation, in order to avoid lepton flavor violating processes such as $\mu\rightarrow e\gamma$. In this way, the Lagrangian reads as
\begin{eqnarray}\label{lag1}
 \mathcal{L} =  \bar{\psi}(\slashed{\partial} + m_\psi)\psi +\left(g_{\psi}S\bar{\psi} e_R + h.c.\right) - V\left(H,S \right),
\end{eqnarray}
where the potential is given by
\begin{eqnarray}\label{lag2}
 V(H,S) &=&\mu_H^2|H|^2 +\lambda_H|H|^4 + \mu_S^2|S|^2 + \lambda_S|S|^4 \nonumber \\ 
 &+& \lambda_{SH}|S|^2|H|^2  + \frac{\mu_3}{2}(S^3 + S^{\dagger 3}).
\end{eqnarray}
The parameters $m_\psi, g_\psi$ and $\mu_3$ can be made real by field redefinitions. The singlet scalar does not acquire vacuum expectation value (vev), and after EWSB (i.e. $\mu_H^2 < 0$) the masses of the scalars become 
\begin{eqnarray}
 m_h^2 &=& -2\mu_H^2, \\
 m_S^2 &=& \mu_S^2 + \lambda_{SH}v_H^2/2,
 \end{eqnarray}
where we recognize $m_h = 125$ GeV as the SM Higgs boson, and $v_H = 246$ GeV. The relevant parameter space of the model is then
\begin{eqnarray}
 \{m_\psi, m_S, g_\psi, \lambda_{SH}, \mu_3\}.
\end{eqnarray}
The $Z_3$ symmetry ensures the stability of the DM particle provided $m_S < 2m_\psi$. Since we are dealing with a complex scalar field, the DM candidate is not self-conjugate and $S$ and $S^*$ are the DM particles.

\section{Constraints}\label{constrains}
In this section we review the most relevant constraints on the model. We focus our attention to constraints that are relevant for new physics in the mass range GeV-TeV, and couplings bigger than $\mathcal{O}(10^{-2})$. In most of the constraints we assume $\lambda_{SH} = 0$, in order to emphasize the novelty of the new physics related to the new Yukawa-like interaction in eq.~\ref{lag1} and the cubic interaction in eq.~\ref{lag2}.
\subsection{Theoretical}
We assume that the Yukawa-like coupling $g_\psi$ is positive (negative values do not change the physical amplitudes) and that perturbativity sets $g_\psi < \sqrt{4\pi}$. Additionally, the stability of the EW vacuum state that
\begin{eqnarray}\label{pert}
 \lambda_H > 0,\quad \lambda_S > 0,\quad \lambda_{SH} > -\sqrt{\frac{2}{3}\lambda_H\lambda_S}.
\end{eqnarray}
Assuming perturbativity, we also assume that the maximum values of each coupling in \ref{pert} is at most 4$\pi$. 

On the other hand, the value of $\mu_3$ can not be too large because it enters in conflicts with the $Z_3$-breaking extrema, threatening the vacuum stability of the scalar potential. In \cite{Belanger:2012zr} was found that $\mu_3$ must fulfill the following relation  
\begin{eqnarray}\label{bound}
 \text{max}(\mu_3) \approx 2\sqrt{2}\sqrt{\frac{\lambda_S}{\delta}}m_S,
\end{eqnarray}
with $\delta$ a dimensionless parameter which regulates whether the SM vacuum is or not a global minimum. Here we simply take $\delta=2$ (an absolute stable vacuum), and considering the maximum perturbative value for $\lambda_S$, in this work we will have that $\text{max}(\mu_3) \approx 2m_S$.

\subsection{Collider}\label{colliderbounds}
At hadron colliders, pair production of $\psi$ via Drell-Yan processes is the main production mechanism. Since we focus on the freeze-out of the complex scalars, we scan the Yukawa-like coupling in the range $g_\psi \sim \mathcal{O}(10^{-2} - \pi)$, which makes all the subsequent decays of $\psi$ short-lived, then collider bounds do not depend on $g_\psi$, but only on the pair $(m_\psi,m_S)$. We use the limit projections obtained in \cite{Bai:2014osa} for the $Z_2$ version of the model in eq.~\ref{lag1} (i.e., no cubic term for the complex scalar, which at collider level results to be unimportant). In \cite{Bai:2014osa}, signal and backgrounds events for proton-proton collisions at 14 TeV LHC and $\mathcal{L} =$ 100 fb$^{-1}$ were simulated for cross sections at leading-order using MadGraph code \cite{Alwall:2011uj}, with the corresponding events showered and hadronized with Pythia \cite{Sjostrand:2007gs} and the jets clustered using of PGS, performing the corresponding exclusion of parameter space at 90\% C.L. These bounds are in agreement to the recent results obtained in \cite{Guedes:2021oqx}.

Additionally, we include bounds from compressed spectra, i.e. $m_\psi \approx m_S$, taken from \cite{Athron:2021iuf}. These bounds are constructed by simulated proton-proton collisions at the LHC at $\sqrt{s} = $13 TeV, with a real scalar plus a VLL with $Y=-1$, and both $SU(2)_L$ singlets. Even when these limits are obtained for the new physics coupled to muons of the SM, the bounds end up being similar to our case, in which the new physics couples only to $e^\pm$ \cite{Kawamura:2020qxo}. 

If the Higgs portal is open and $m_h > 2m_S$, then the Higgs decay into $SS(S^*S^*)$ contributes to the decay width $\Gamma_h$ of the Higgs given by \cite{Cline:2013gha}
\begin{eqnarray}
 \Gamma_\text{inv} = 2\times \frac{\lambda_{SH}^2 v_H^2}{32\pi m_h}\sqrt{1 - \frac{4m_S^2}{m_h^2}},
\end{eqnarray}
with the factor 2 in the right side account for the Higgs decay into both $SS$ and $S^*S^*$. Upper bounds on Br$(h\rightarrow$ inv) = $\Gamma_\text{inv.}/(\Gamma_\text{SM} + \Gamma_\text{inv.})$, with $\Gamma_\text{SM} = 4.07$ MeV \cite{CMS:2018yfx}, set that Br$(h\rightarrow$ inv.) $< 0.19$ at 95\% C.L. \cite{CMS:2018yfx}, implying that $\lambda_{SH} \lesssim 10^{-2}$. Finally, we have checked that the new physics contribution to the decay $Z\rightarrow e^+e^-$ at one-loop results to be completely negligible to constraint the parameter space of the model.
\begin{figure}[t!]
\centering
\begin{tikzpicture}[line width=1.0 pt, scale=0.5]

\begin{scope}[shift={(0,0)}]
	\draw[scalarnoarrow](-2.5,1) -- (-1,0);
	\draw[scalarnoarrow](-2.5,-1) -- (-1,0);
	\draw[scalarnoarrow](-1,0) -- (1,0);
	\draw[fermion](1,0) -- (2.5,1);
	\draw[fermionbar](1,0) -- (2.5,-1);
    \node at (-3,1.0) {$S$};
	\node at (-3,-1.0) {$S$};
    \node at (-0.1,0.46) {$S^*$};
	\node at (3,1.0) {$e^-$};
    \node at (3,-1.0) {$\psi^+$};
    
    \node at (-0,-2.5) {$(a)$};
\end{scope}

\begin{scope}[shift={(8,0)}]
 \draw[scalarnoarrow](-3,1) -- (-1,1);
	\draw[scalarnoarrow](-3,-1) -- (-1,-1);
	\draw[fermionbar](-1,1) -- (-1,-1);
	\draw[fermion](-1,1) -- (1,1);
	\draw[fermionbar](-1,-1) -- (1,-1);
    \node at (-3.6,1.0) {$S$};
	\node at (-3.6,-1.0) {$S^*$};
    \node at (-0.5,0) {$\psi$};
	\node at (1.6,1) {$e^-$};
    \node at (1.6,-1) {$e^+$}; 
    \node at (-0.8,-2.5) {$(b)$};
\end{scope}

\begin{scope}[shift={(0,-5)}]
 \draw[scalarnoarrow](-3,1) -- (-1,1);
	\draw[fermion](-3,-1) -- (-1,-1);
	\draw[fermionbar](-1,1) -- (-1,-1);
	\draw[fermion](-1,1) -- (1,1);
	\draw[vector](-1,-1) -- (1,-1);
    \node at (-3.6,1.0) {$S$};
	\node at (-3.6,-1.0) {$\psi^-$};
    \node at (-0.5,0) {$\psi$};
	\node at (1.6,1) {$e^-$};
    \node at (1.7,-1) {$A, Z$}; 
    \node at (-0.8,-2.5) {$(c)$};
\end{scope}

\begin{scope}[shift={(7,-5)}]
	\draw[scalarnoarrow](-2.5,1) -- (-1,0);
	\draw[fermion](-2.5,-1) -- (-1,0);
	\draw[fermion](-1,0) -- (1,0);
	\draw[fermion](1,0) -- (2.5,1);
	\draw[vector](1,0) -- (2.5,-1);
    \node at (-3,1.0) {$S$};
	\node at (-3,-1.0) {$\psi^-$};
    \node at (0.2,0.7) {$e^-$};
	\node at (3,1.0) {$e^-$};
    \node at (3.2,-1.0) {$A, Z$};
    
    \node at (-0,-2.5) {$(d)$};
\end{scope}
\end{tikzpicture}
\caption{\textit{Leading processes producing the relic abundance of $S$ at freeze-out. Radiative corrections, Higgs portal contributions, and the corresponding CP transformed process are not shown.}}\label{diagrams}
\end{figure}
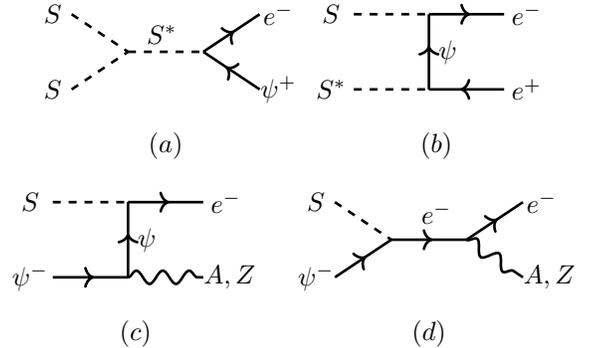
\begin{figure}[t!]
\centering
\includegraphics[width=0.4\textwidth]{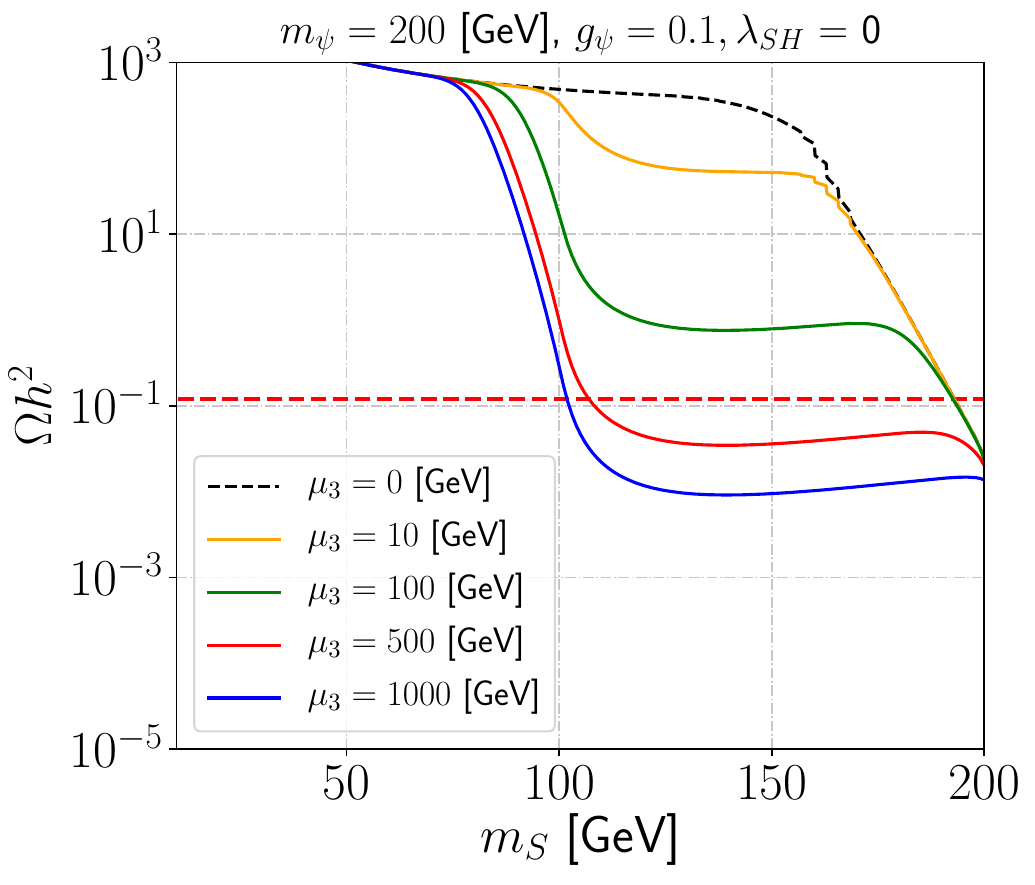}
\includegraphics[width=0.4\textwidth]{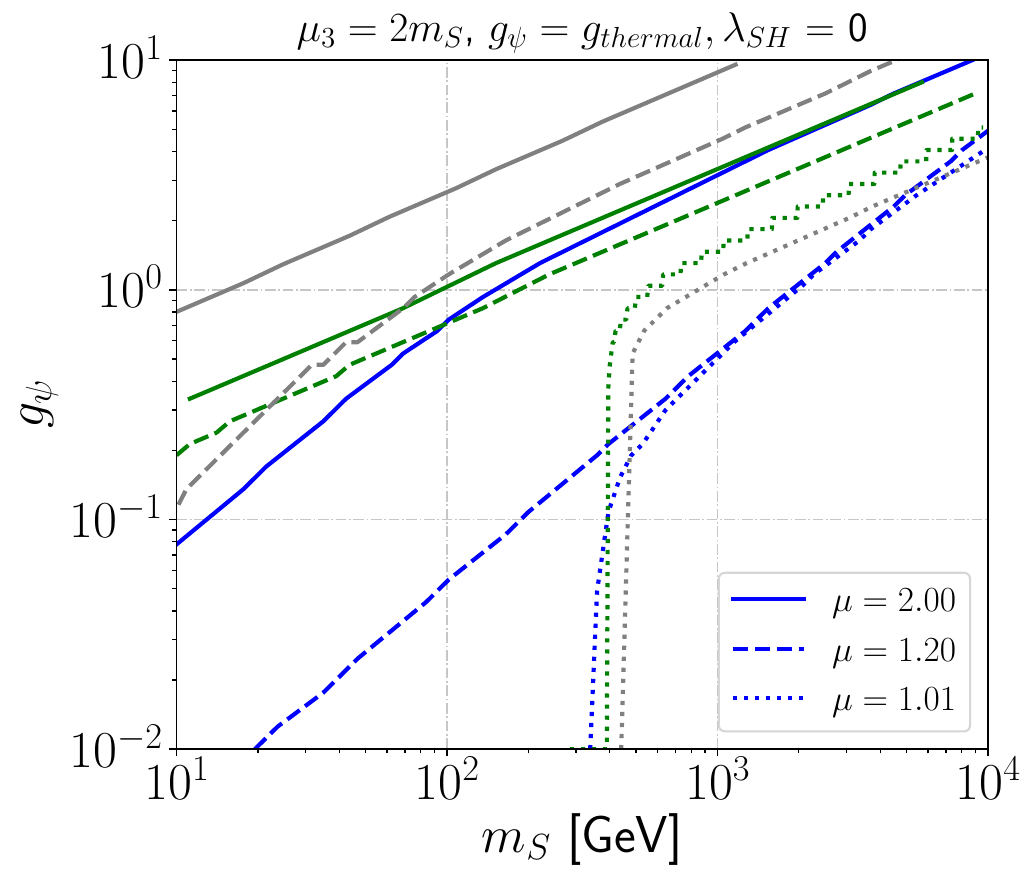}
\caption{\textit{(above) Micromegas predictions for the relic density abundance of $S(S^*)$ considering the values of the parameters shown in the plot. The red dotted horizontal lines correspond to the measured relic abundance $\Omega_c h^2 = 0.12$. (below) Yukawa coupling $g_\psi$ leading to the observed DM relic density in the $Z2R$ model (grey curves), in the $Z2C$ model (green curves), i.e. $\mu_3 = 0$, and in the present $Z_3$ model (blue curves) with $\mu_3 = 2m_S$, for different values of $\mu = m_\psi/m_S$.}} 
\label{pred1}
\end{figure}
\subsection{Relic Abundance}\label{relicab}
In the calculation of the relic abundance, we have implemented the model on LanHEP \cite{Semenov:2002jw} and on MicrOMEGAS 5.2.7.a code \cite{Belanger:2013ywg}. We focus on the freeze-out of $S(S^*)$ particles, with the relic density determined by the processes presented in Fig.~\ref{diagrams}, plus Higgs portal contributions and higher order corrections. One of the novelties of this work is the presence and relevance of the diagram (a) of Fig.~\ref{diagrams}, with $\braket{\sigma_{SS\rightarrow e\psi} v} \propto \mu_3^2g_\psi^2$ (equivalent for its CP-conjugate process) occurring in the s-wave. To exemplify the influence of this process in the calculation of the relic abundance, in Fig.~\ref{pred1}(\textit{above}) we show the values of the relic density calculation as a function of $m_S$ for different values of $\mu_3$, keeping the rest of the parameters shown at the top of the plot. Notice that deviations from $\mu_3=0$ curve (dashed black) for $m_S \lesssim m_\psi$, where the process (a) becomes dominant, changing the relic abundance up to several orders of magnitude. For $m_S\lesssim m_\psi/2$ annihilations of the type $SS(S^*S^*)$ can still be effective due to the thermal tail. In this way, provided that $m_\psi/2 \lesssim m_S \lesssim m_\psi$, the influence of the new physics dictated by diagram (a) may produce strong effects in the relic density calculation \footnote{MicrOMEGAS code does not take into account $2\rightarrow 3$ processes in the relic density calculation, unless there is a gauge boson radiated in the final state. In the mass range $m_S < m_\psi/2$, we should expect some non-negligible contributions of $2\rightarrow 3$ processes, although this mass regime is not part of our interest in this work.}.

A $Z_2$ version of this model has been studied in the past, consisting of either real or complex singlet scalar with a VLL mediator. In their minimal form, both new fields are $SU(2)_L$ singlets, the VLL with $Y=-1$ and the new physics only coupled to the first family of leptons. We call those models \textit{Z2R} and \textit{Z2C}, respectively (references of these type of scenarios can be found in Sec.~\ref{int}). The resulting Lagrangian is similar to eq.~\ref{lag1} without the $\mu_3$ term. In Fig.~\ref{pred1}(\textit{below}) we compare the deviations in the thermal Yukawa couplings in our model (blue curves) for $\mu_3 = 2m_S$ with respect to the results obtained in \textit{Z2R} (grey lines) and \textit{Z2C} (green lines), with the rest of the parameters equally fixed. The new annihilation process from the $\mu_3$ term makes the Yukawa couplings in the $Z_3$ model milder, mostly independent of the mass shift $\mu\equiv m_\psi/m_S$, provided that $m_S \gtrsim m_\psi/2$. The differences between \textit{Z2R} and \textit{Z2C} are due to the fact that their respective DM annihilation cross sections present different suppressions in their partial waves.

The fact that Yukawa couplings tend to decrease in our framework provided that $\mu_3\neq 0$ have interesting effects on collider and dark matter phenomenology.

\subsection{Direct detection}
Direct detection (DD) bounds for DM candidates with masses of $\mathcal{O}(100)$ GeV are set by XENON1T \cite{Aprile_2018}. In absence of Higgs portal, DM-nucleon interactions via the VLL portal start at one-loop through the dimension-six charge-radius operator $\mathcal{L} \sim C \partial_\mu S^* \partial_\nu S F^{\mu\nu}$ \cite{Chang:2014tea, Bai:2014osa, Kawamura:2020qxo}, with the total cross section for the DM-nucleon system given by
\begin{eqnarray}
 \sigma_{SN} = \frac{Z^2e^2C^2(m_e, m_\psi)\mu_{SN}^2}{8\pi A^2}, 
\end{eqnarray}
with $Z$ the atomic number, $e$ the electric charge, $\mu_{SN}\equiv m_Sm_N/(m_S + m_N)$, where $m_N = 0.94$ GeV is the nucleon mass, and $C(m_e,m_\psi)$ in the limit $m_e\ll m_\psi$ is given by
\begin{align*}
 C(m_e,m_\psi) = -\frac{g_\psi^2 e}{16\pi^2 m_\psi^2}\left(1 + \frac{2}{3}\log\left(\frac{m_e^2}{m_\psi^2}\right)\right].
\end{align*}
In the case of Higgs portal interactions, the spin-independent cross section is present at tree-level. If the scalar DM gives the correct relic abundance, we have then the Higgs portal coupling must satisfy \cite{Cline:2013gha}
\begin{eqnarray}
 \lambda_{SH} \leq \left(\frac{4\pi m_h^4 m_S^2 \sigma_{Xe}}{f_N\mu_{SN}^2m_N^2}\right)^{1/2} ,
\end{eqnarray}
with $\sigma_\text{Xe}$ the upper bounds given by XENON1T, and the Higgs-nucleon coupling is given by $f_N \approx 0.3$.
\subsection{Indirect detection}
In the following we detail gamma-ray and positron predictions in the $Z_3$ model, with a cross comparation with \textit{Z2R} and \textit{Z2C} models. On the other hand, anti-protons come from $SS^*\rightarrow e^+e^-Z$ process via the decay and hadronization of the $Z$ boson. However, bounds from anti-protons are expected to be even smaller than what is obtained in \textit{Z2R} and \textit{Z2C} \cite{Giacchino:2013bta, Ibarra:2014qma}, since once $\mu_3$ takes sizable values, the values of the Yukawa couplings tend to decrease (see Sec.~\ref{relicab}), in turn decreasing the annihilation of the aforementioned process. Therefore, we do not expect sizable bounds on this lepto-philic realization of the $Z_3$ scenario from anti-protons measurements.

\subsubsection*{Gamma-rays}
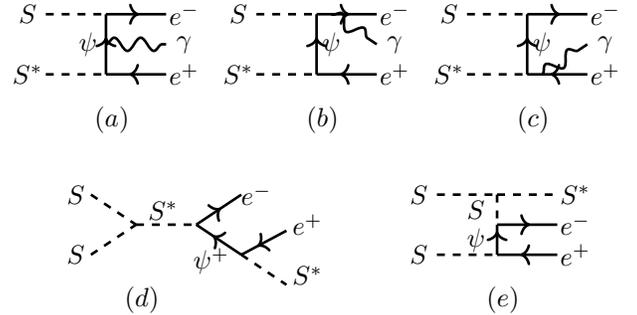
\begin{figure}[t!]
\centering
\begin{tikzpicture}[line width=1.0 pt, scale=0.4]
\begin{scope}[shift={(-2,4)}]
 \draw[scalarnoarrow](-3,1) -- (-1,1);
	\draw[scalarnoarrow](-3,-1) -- (-1,-1);
	\draw[fermionbar](-1,1) -- (-1,-1);
	\draw[fermion](-1,1) -- (1,1);
	\draw[vector](-1,0) -- (1,0);
	\draw[fermionbar](-1,-1) -- (1,-1);
    \node at (-3.6,1.0) {$S$};
	\node at (-3.6,-1.0) {$S^*$};
    \node at (-1.6,0) {$\psi$};
	\node at (1.6,1) {$e^-$};
	\node at (1.6,0) {$\gamma$};
    \node at (1.6,-1) {$e^+$}; 
    \node at (-0.8,-2.5) {$(a)$};
\end{scope}

\begin{scope}[shift={(5,4)}]
 \draw[scalarnoarrow](-3,1) -- (-1,1);
	\draw[scalarnoarrow](-3,-1) -- (-1,-1);
	\draw[fermionbar](-1,1) -- (-1,-1);
	\draw[fermion](-1,1) -- (1,1);
	\draw[vector](-0.5,1) -- (1,0);
	\draw[fermionbar](-1,-1) -- (1,-1);
    \node at (-3.6,1.0) {$S$};
	\node at (-3.6,-1.0) {$S^*$};
    \node at (-0.5,0) {$\psi$};
	\node at (1.6,1) {$e^-$};
	\node at (1.6,0) {$\gamma$};
    \node at (1.6,-1) {$e^+$}; 
    \node at (-0.8,-2.5) {$(b)$};
\end{scope}

\begin{scope}[shift={(12,4)}]
 \draw[scalarnoarrow](-3,1) -- (-1,1);
	\draw[scalarnoarrow](-3,-1) -- (-1,-1);
	\draw[fermionbar](-1,1) -- (-1,-1);
	\draw[fermion](-1,1) -- (1,1);
	\draw[vector](-0.5,-1) -- (1,0);
	\draw[fermionbar](-1,-1) -- (1,-1);
    \node at (-3.6,1.0) {$S$};
	\node at (-3.6,-1.0) {$S^*$};
    \node at (-0.5,0) {$\psi$};
	\node at (1.6,1) {$e^-$};
	\node at (1.6,0) {$\gamma$};
    \node at (1.6,-1) {$e^+$}; 
    \node at (-0.8,-2.5) {$(c)$};
\end{scope}

\begin{scope}[shift={(-1,-2)}]
	\draw[scalarnoarrow](-2.5,1) -- (-1,0);
	\draw[scalarnoarrow](-2.5,-1) -- (-1,0);
	\draw[scalarnoarrow](-1,0) -- (1,0);
	\draw[fermion](1,0) -- (2.5,1);
	\draw[fermionbar](1,0) -- (2.5,-1);
	\draw[fermionbar](2.5,-1) -- (4,-0.2);
	\draw[scalarnoarrow](2.5,-1) -- (4,-2);
    \node at (-3,1.0) {$S$};
	\node at (-3,-1.0) {$S$};
    \node at (-0.1,0.46) {$S^*$};
	\node at (3,1.0) {$e^-$};
    \node at (1.5,-1.2) {$\psi^+$};
    \node at (4.7,0) {$e^+$};
    \node at (4.7,-1.7) {$S^*$};
    \node at (-0.8,-2.5) {$(d)$};
\end{scope}

\begin{scope}[shift={(11,-2)}]
 \draw[scalarnoarrow](-3,1) -- (-1,1);
	\draw[scalarnoarrow](-3,-1) -- (-1,-1);
	\draw[fermionbar](-1,0) -- (-1,-1);
	\draw[scalarnoarrow](-1,1) -- (1,1);
	\draw[scalarnoarrow](-1,1) -- (-1,0);
	\draw[fermionbar](-1,-1) -- (1,-1);
	\draw[fermion](-1,0) -- (1,0);
    \node at (-3.6,1.0) {$S$};
	\node at (-3.6,-1.0) {$S$};
	\node at (-1.7,0.5) {$S$};
    \node at (-1.7,-0.5) {$\psi$};
	\node at (1.6,1) {$S^*$};
	\node at (1.6,0) {$e^-$}; 
    \node at (1.6,-1) {$e^+$}; 
    \node at (-0.8,-2.5) {$(e)$};
\end{scope}
\end{tikzpicture}
\caption{\textit{Diagrams (a), (b) and (c) are Bremsstrahlung processes appearing from the $SS^*$ annihilation. The diagrams (d) and (e) correspond to the annihilation $SS$ (equivalent for $S^*S^*$), giving rise to the box-shape spectra for electrons and positrons.}}\label{diagrams2}
\end{figure}

\begin{figure}[t!]
\centering
\includegraphics[width=0.4\textwidth]{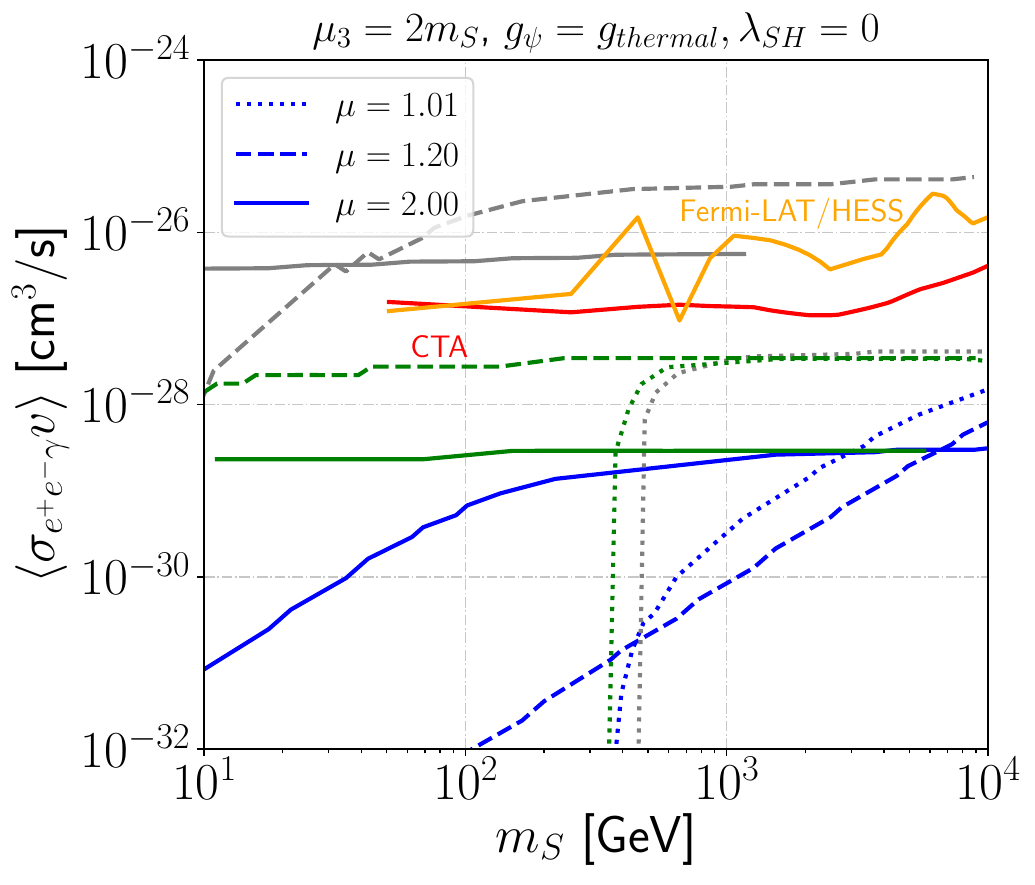}
\includegraphics[width=0.4\textwidth]{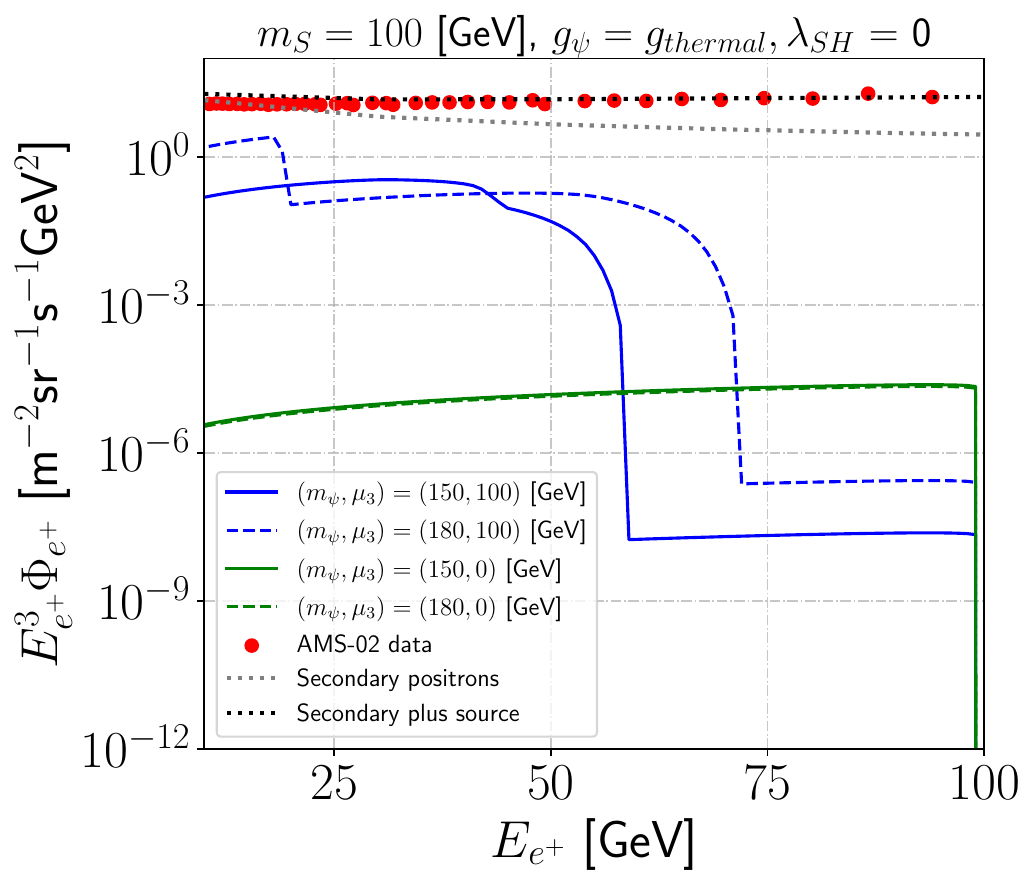}
\caption{\textit{(above) Average annihilation cross section $\braket{\sigma_{e^+e^-\gamma} v}$ as function of the DM mass, with the blue curves representing the results obtained in the $Z_3$ model, and the grey(green) ones obtained in the \textit{Z2R}(\textit{Z2C}) model. The Yukawa coupling $g_\psi$ has been fixed to the value leading to the observed relic abundance. The orange and red solid lines represent bounds from Fermi-LAT/HESS and expected sensitivity for CTA, respectively. (below) Positron flux as a function of the positron energy for the parameters shown in the figure. For comparision, we have added AMS-02 data \cite{PhysRevLett.122.041102}, along with the secondary and astrophysical source positrons assumption performed in \cite{Ibarra:2013zia}.}} 
\label{sigvA}
\end{figure}

Since we are interested in the mass regime $m_\psi/2 \lesssim m_S < m_\psi$, this translate into $\mu \ll 10$, then one-loop annihilations $SS^*\rightarrow \gamma\gamma, \gamma Z$ are negligible in comparison to $SS^*\rightarrow e^+e^-\gamma$ radiative process \cite{Ibarra:2014qma} (see Fig.~\ref{diagrams2} upper row). These latter processes $2\rightarrow 3$ are the leading gamma-ray sources in the parameter space relevant for the present analysis here. Additionally, as it is shown in Fig.~\ref{pred1}(\textit{below}), in this mass regime we expect smaller Yukawa couplings with respect to the ones obtained in \textit{Z2R} and \textit{Z2C}, then decreasing gamma-ray signals even more than \textit{Z2C} model, then indirect searches putting none constraint on the parameter space of the $Z_3$ model. To exemplify this fact, in Fig.~\ref{sigvA}\textit{(above)} we show the corresponding values for the average cross section today into gamma-rays coming from $2\rightarrow 3$ process for different $\mu$ values, fixing $g_\psi$ to its thermal value, and $\mu_3 = 2m_S$. The blue lines, representing the predictions of the $Z_3$ model, tend to be reduced in comparison with the ones predicted by \textit{Z2R} (grey lines) and \textit{Z2C} (green lines), then being some orders of magnitudes below Fermi-LAT and CTA constraints \cite{Garny:2013ama}\footnote{The upper bounds where constructed for Majorana DM in a scalar portal, and they can be used in our context provided that Bremsstrahlung process dominates over the one loop processes $2\rightarrow 2$ \cite{Garny:2013ama}. This is exactly our case in which we have $\mu\lesssim 2$.}. Notice that $\mu_3 = 2m_S$ is just an arbitrary election, and smaller $\mu_3$ will force to increase $g_\psi$ to maintain the correct relic abundance, automatically increasing the gamma-ray signal, but never above the predictions of \textit{Z2C}. 

The predictions in Fig.~\ref{sigvA}\textit{(above)} for the $Z_3$ model are understood by the fact that the relic density is obtained mainly by $\braket{\sigma_{SS\rightarrow e\psi}v}$, then since $\Omega_S h^2 \sim Y_0 m_S$ with $Y_0 \sim \braket{\sigma_{SS\rightarrow e\psi}v}^{-1} \propto m_S^2/g_\psi^2$, we have $\Omega_S h^2 \sim m_S^3/g_\psi^2$, this is, as $m_S$ increases, $g_\psi$ must increase too to keep the correct relic abundance, which is translated into the fact that higher DM masses produce stronger gamma-ray lines signals.
\subsubsection*{Positrons}
In the annihilation of DM today, the following processes contribute to the positron fluxes:
\begin{enumerate}\label{enumerate}
 \item $SS^*\rightarrow e^+e^-$,
 \item $SS^*\rightarrow e^+e^-\gamma(Z)$,
 \item $SS\rightarrow e^+e^-S^*$ (and its CP-conjugate process).
\end{enumerate}
The process (1) annihilates in the $p$-wave (similar to Majorana DM), then velocity suppressed. Process (2) proceeds in the $s$-wave but suppressed by $\alpha_\text{em}$, and additional pair of $e^\pm$ may be emitted from $\gamma/Z$. For $m_\psi/2 \lesssim m_S < m_\psi$, the process (3) results to be the dominant one, increasing $e^\pm$ fluxes by several orders of magnitude in certain positron energy range in comparison to the other two processes (provided $\mu_3 \neq 0$), since the diagram of Fig.~\ref{diagrams2}(\textit{d}) proceeds in the $s$-wave, without helicity nor electromagnetic suppressions. 

As an example of the enhancement of the positron flux measured at earth, in Fig.~\ref{sigvA}(\textit{below}) we show the energy flux of $e^+$ obtained with MicrOMEGAS, for the model parameter values specified in the plot\footnote{MicrOMEGAS code takes into account cosmic-rays propagation effects solving the diffusion-loss equation keeping space diffusion and energy losses. The parameters taken into account are MED propagation parameters: diffusion coefficient $\mathcal{K}_0 = 0.0112$ kpc$^2/$Myr, $\delta = 0.7$ and the energy loss coefficient $b_0 = 10^{-16}$ s$^{-1}$.}. The blue lines consider $\mu_3 = 100$ GeV and the green curves $\mu_3 = 0$, with $m_S = 100$ GeV, keeping $g_\psi$ to its thermal value and $m_\psi$ indicated in each curve. As it is shown in the plot, the cases with $\mu_3 = 100$ GeV (blue curves) manifest high positron fluxes in certain positron energies in comparison to the case in which $\mu_3 = 0$ (green curves). The former fluxes correspond to $S + S\rightarrow e^- + \psi^+(\rightarrow e^+ + S^*)$, presenting a \textit{box-shaped} spectra, and from the two-body annihilation $S^*S^*\rightarrow e^+ + \psi^-$, with a sharp peak energy (kinematic details and normalized spectra are shown in Appx.~\ref{app}). By kinematics, high energy positrons are not able to be produced by these latter processes, and the radiative $SS^* \rightarrow e^+ e^- \gamma$ process enter to dominate, as can be noted for instance for $E_{e^+} \gtrsim 55(65)$ GeV for $m_\psi = 150(180)$ GeV. The sharped edges of the blue curves around $E_{e^+}\sim 55$ and $65$ GeV come from the fact that MicrOMEGAS does not consider all the relevant channels that could contribute to the positron fluxes at all energies. We have checked with CalcHEP (see Appx.~\ref{app}) that an off-shell $\psi$ in diagram (d) of Fig.~\ref{diagrams2} makes the spectra smoother\footnote{We acknowledge the referee for pointing out this interesting issue. Internal communication with the authors of MicrOMEGAS mentioned us that this type of contributions are going to be considered in the new versions of the code.}. Similarly, positron fluxes for $2m_S < m_\psi$ results to be suppressed by the off-shellness of diagram (d).

In order to constrain our scenario with positron measurements, we make use of model independent-bounds on the annihilation $\braket{\sigma_{\text{DMDM}\rightarrow ee} v}$ obtained in \cite{Ibarra:2013zia} for the positron flux data from AMS-02. These bounds were obtained assuming a power-law background fit of AMS-02 data based on a contribution from secondary positrons produced in cosmic ray spallations and a possibly contribution from astrophysical sources, and a signal contribution coming from the DM annihilation. The only parameter space points susceptible to be constrained in our $Z_3$ scenario are those with high positron fluxes, i.e. $2m_S > m_\psi$ and $\mu_3\neq 0$, therefore our criteria to evade AMS-02 bounds is $\braket{\sigma_{\text{SS}\rightarrow e\psi} v} < \braket{\sigma_{\text{DMDM}\rightarrow ee} v}$. 

\section{Scan Results}\label{results}
In this section we show the viable parameter space of the model for some values of $\mu_3$ after imposing all the constraints previously described. We keep the Higgs portal equal to zero, and in the next section we discuss about the effects of this portal on the results presented here. In Fig.~\ref{all1} we show the resulting parameter space for $\mu_3 = 0, m_S/2, 2m_S$ (from top row to bottom row, respectively), keeping the LHC bounds as the black contour (LHC 14 TeV projection), the orange region (compressed LHC 13 TeV spectra) and the cyan for VLL searches (see Sec.~\ref{colliderbounds}), green regions in each plot are the parameter space where $m_S > m_\psi$, then $S$ being unstable, and the purple region showing the parameter space in which co-annihilations contribute with more than $20\%$ to the relic generation. In the first column of plots we consider the parameter space after imposing perturbativity and the correct relic abundance. In the second column we present the parameter space left after applying AMS-02 bounds. The third column shows the parameter space left after imposing XENON1T.  

Considering the results of the first raw of Fig.~\ref{all1} with $\mu_3 = 0$ (i.e. \textit{Z2C}), the correct relic abundance is obtained for $g_\psi \approx \mathcal{O}(1 - 10)$, although for smaller $m_S$ the coupling $g_\psi$ tends to increase outside the perturbative limit (first plot to the left). As it expected from \textit{Z2C}, AMS-02 bounds do not exclude none of the parameter space points (middle plot), but XENON1T ruled out most of the parameter space up to 1 TeV (plot to the right), with a small fraction of points in the purple region $m_S \approx m_\psi$ surviving direct detection. This result is agreement with \cite{Bai:2014osa, Chang:2014tea}.

For $\mu_3 \neq 0$, second and third raws of Fig.~\ref{all1}, notoriously smaller $g_\psi$ are obtained in the region $2m_S > m_\psi$. In these cases, AMS-02 bounds become relevant in the low mass region fulfilling $2m_S > m_\psi$, excluding DM masses between 100 - 200 GeV, approximately. As $\mu_3 = 2m_S$, XENON1T bounds do not rule out the remaining parameter space in the region $2m_S > m_\psi$, whereas as $\mu_3 = m_S/2$ most of the parameter space is under tension under this constraint, although some parameter space still survives. We have checked that XENONnT projections \cite{XENON:2020kmp} ruled out all the remaining points in the case of $\mu_3 = m_S/2$, whereas for $\mu_3 = 2m_S$ most of the parameter space left by XENON1T survives. 

\begin{figure}[t!]
\centering
\includegraphics[width=0.45\textwidth]{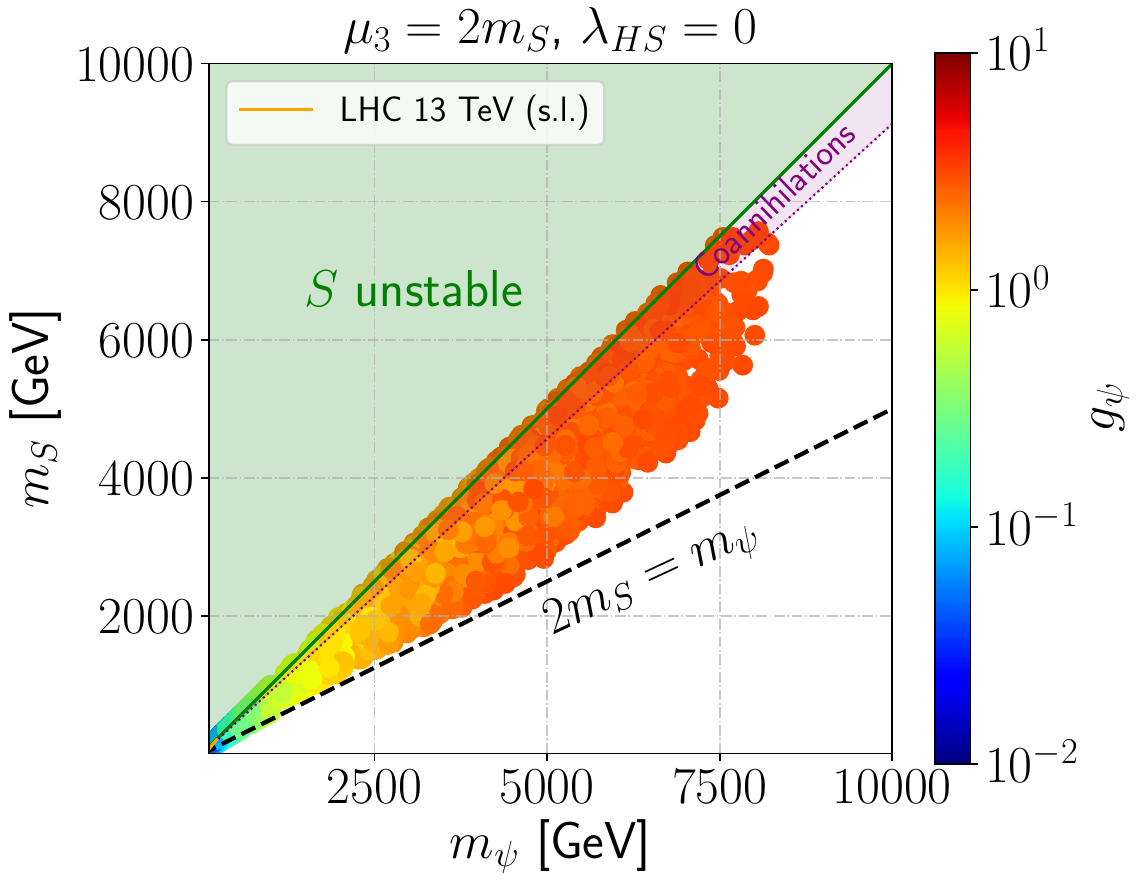}
\caption{\textit{High mass regime considering fulfilling perturbativity and correct relic abundance, and XENONnT and CTA bound projections.}} 
\label{highmass}
\end{figure}

In resume, a parameter space opens up below the TeV region passing the strongest constraints and presenting high sensitivity to experimental searches. As a mode of complementarity, in Fig.~\ref{highmass} we show the prospects of having DM with masses up to 10 TeV, where we have included XENONnT and CTA bounds, considering $g_\psi < \sqrt{4\pi}$. Notice that there is still a vast parameter space left after the imposition of the strongest constraints.

\section{Discussion and Conclusions}\label{conc}
In its simplest form, we have explored a novel model containing two new fields ($S, \psi)$, both transforming under a $Z_3$ symmetry. As $m_S < m_\psi$, $S$ is stable becoming a DM candidate. The model allows a cubic interaction term for $S(S^*)$ leading to the appearance of a new scattering processes relevant for the calculation of the relic abundance and indirect detection, requiring milder Yukawa-like coupling unlike to previous similar constructions based on the $Z_2$ symmetry. This fact \textit{a priori} makes that the model be perturbative at higher energies than \textit{Z2R} and \textit{Z2C} models. At the phenomenological level the characteristic of this model in its simplest form (Yukawa-like coupling only to $e^-/e^+$) are remarkable, due to the fact that some regions of the parameter space not only survive near the EW scale up to several TeV and evading strong direct detection bounds, but also presenting sizable fluxes in positrons with a distinctive box-shaped spectra. 

\begin{figure}[t!]
\centering
\begin{multicols}{3}
\includegraphics[width=0.28\textwidth]{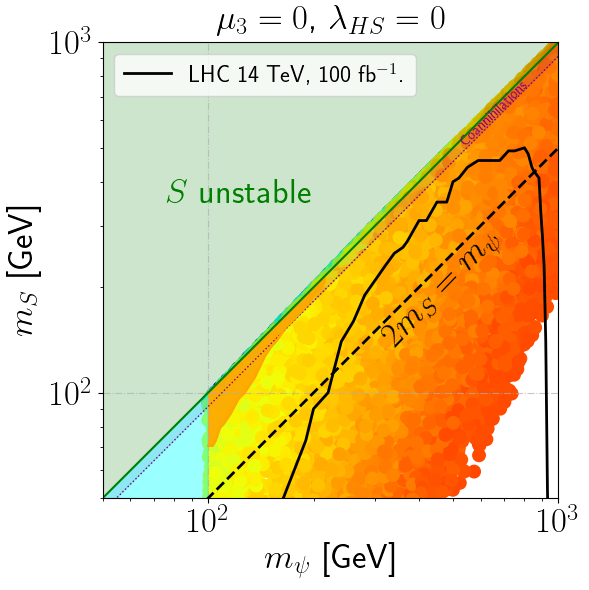}\par
\includegraphics[width=0.26\textwidth]{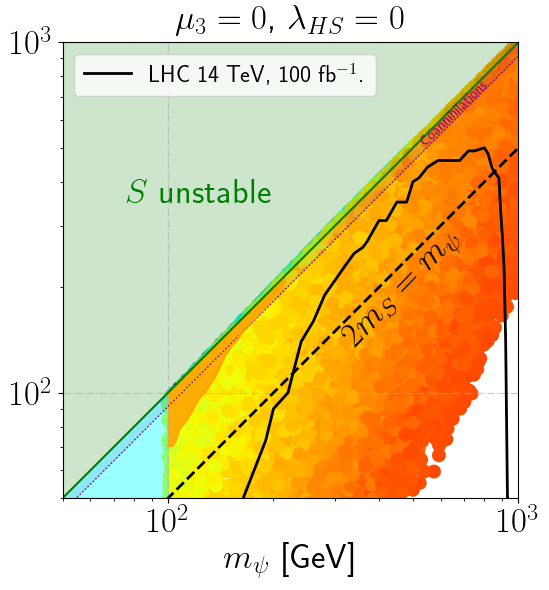}\par
\includegraphics[width=0.32\textwidth]{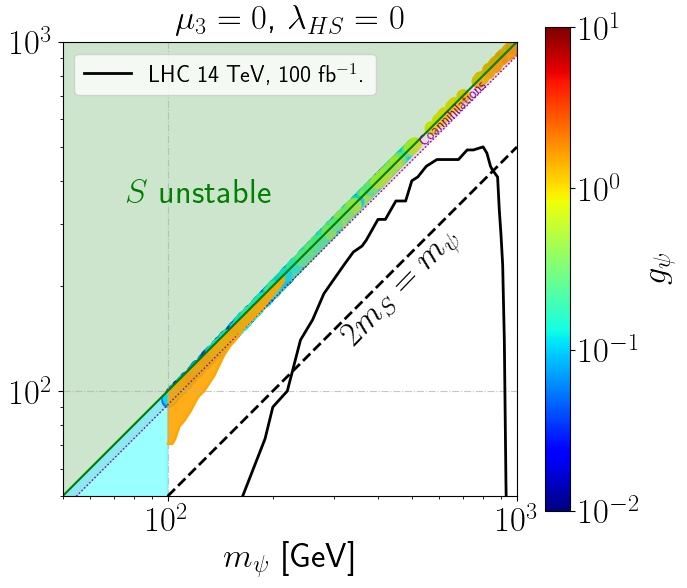}
\end{multicols}
\begin{multicols}{3}
\includegraphics[width=0.28\textwidth]{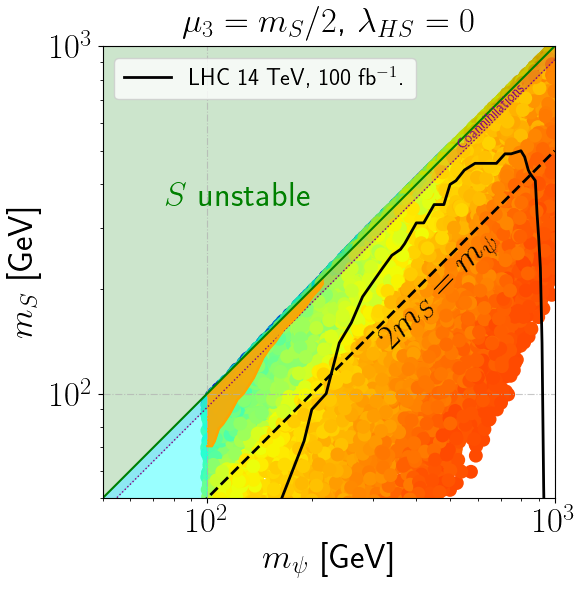}\par
\includegraphics[width=0.26\textwidth]{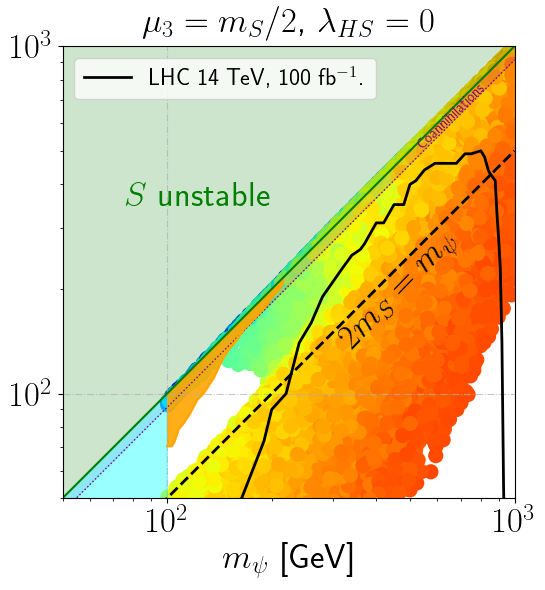}\par
\includegraphics[width=0.32\textwidth]{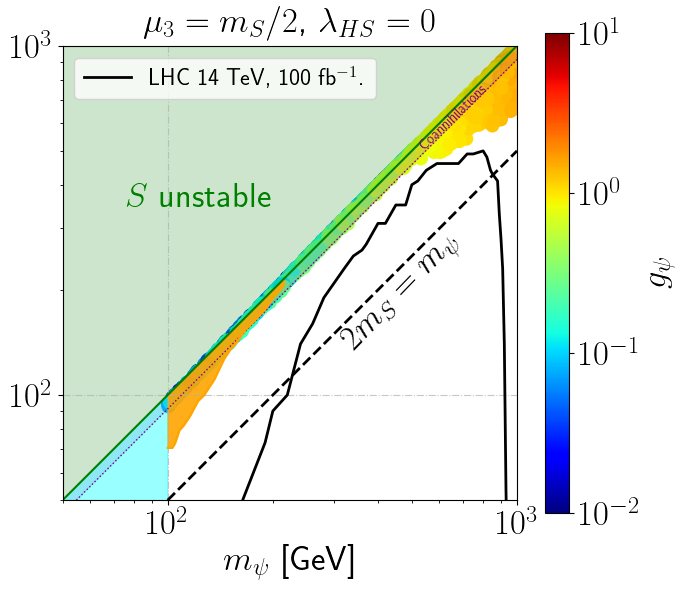}
\end{multicols}
\centering
\begin{multicols}{3}
\includegraphics[width=0.28\textwidth]{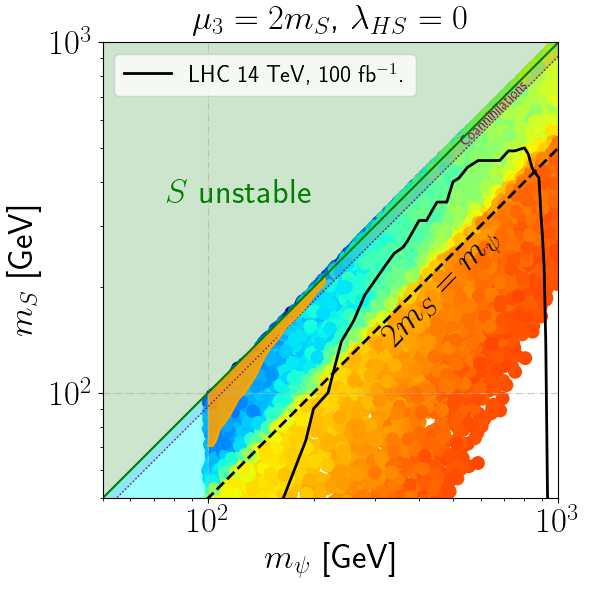}\par
\includegraphics[width=0.26\textwidth]{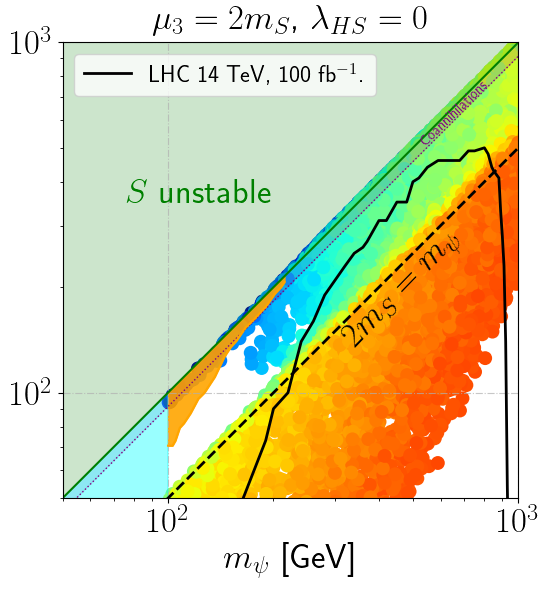}\par
\includegraphics[width=0.32\textwidth]{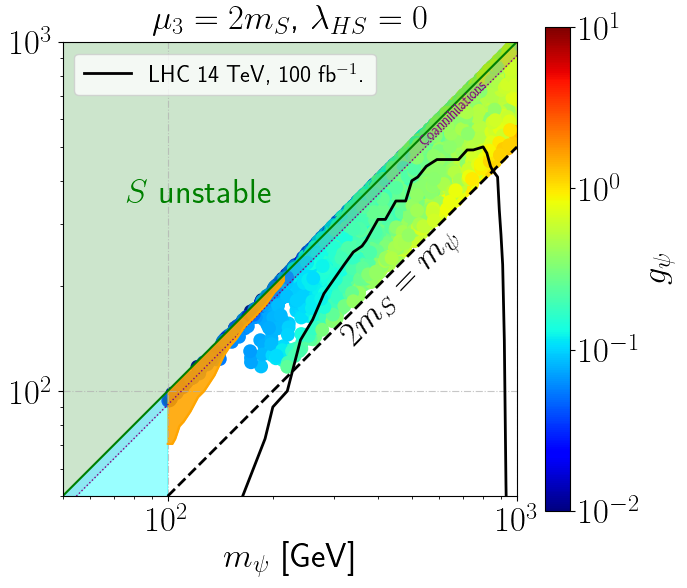}
\end{multicols}
\caption{\textit{Color maps for $g_\psi$ in the mass plane ($m_\psi, m_S$), for $\mu_3 = 0, m_S/2$ and  $2m_S$ (raws from up to down, respectively). The first column of plots keep the correct relic abundance and perturbative bounds on each point, the second one indirect detection bounds (AMS-02 and CTA), and in the third one we add direct detection bounds (XENON1T). The green regions in each plot is forbidden due to the fact that $S$ becomes unstable. The black solid line corresponds to the LHC projection bounds for 14 TeV, the cyan region corresponds to collider constraints for VLL, and the orange region is the exclusion from compressed spectra for LHC at 13 TeV. The dashed line is a reference in which above it the $SS(S^*S^*)$ annihilations start to become efficients, whereas the small purple region represent the place in which co-annihilations start to be effective.}}
\label{all1}
\end{figure}

One of the assumptions made in this work was keeping $\lambda_{SH} = 0$. The consequences of having such portal with sizable values is equivalent to what occurs in these types of scenarios: new channels participating of the relic abundance calculation, new indirect detection signals, and direct detection at tree level. In order to keep the correct relic abundance passing all the previous constraints, the immediate effect of sizable $\lambda_{SH}$ is the reduction of either $g_\psi$ or $\mu_3$ in order to keep the correct relic abundance, then affecting some results found in this work. However, couplings up to order $\lambda_{SH} \sim \mathcal{O}(10^{-2})$ do not deviate the results already found in this work. Just to sketch, consider the parameter space point $(m_\psi, m_S) = (200,130)$ GeV, $g_\psi = 0.1$ and $\mu_3 = 2m_S$. We have checked that it is possible to set $\lambda_{SH} \lesssim 0.01$, evading all the constraints, without generating a significant deviation of the values presented in Fig.~\ref{all1}. 
\onecolumn
\twocolumn

Finally, all the treatment that has been carried out here is assuming a single coupling of the new physics to $e_R$ field. What we have done is for the sake of simplicity and illustration of the possible enhancement of the new annihilation processes of the complex DM, but certainly the coupling of the new sector to other lepton/quark families, simultaneously or not, and/or the dark sector fields in higher SM gauge representations could result in an interesting phenomenology. 

\section{Acknowledgments}
We thank Alejandro Ibarra for helpful discussions and comments, and Alexander Pukhov for helped us with technical things of MicrOMEGAS. B.D.S has been founded by ANID (ex CONICYT) Grant No. 74200120.



\appendix

\section{Electron/Positron kinematics}\label{app}
\begin{figure}[t!]
\centering
\includegraphics[width=0.4\textwidth]{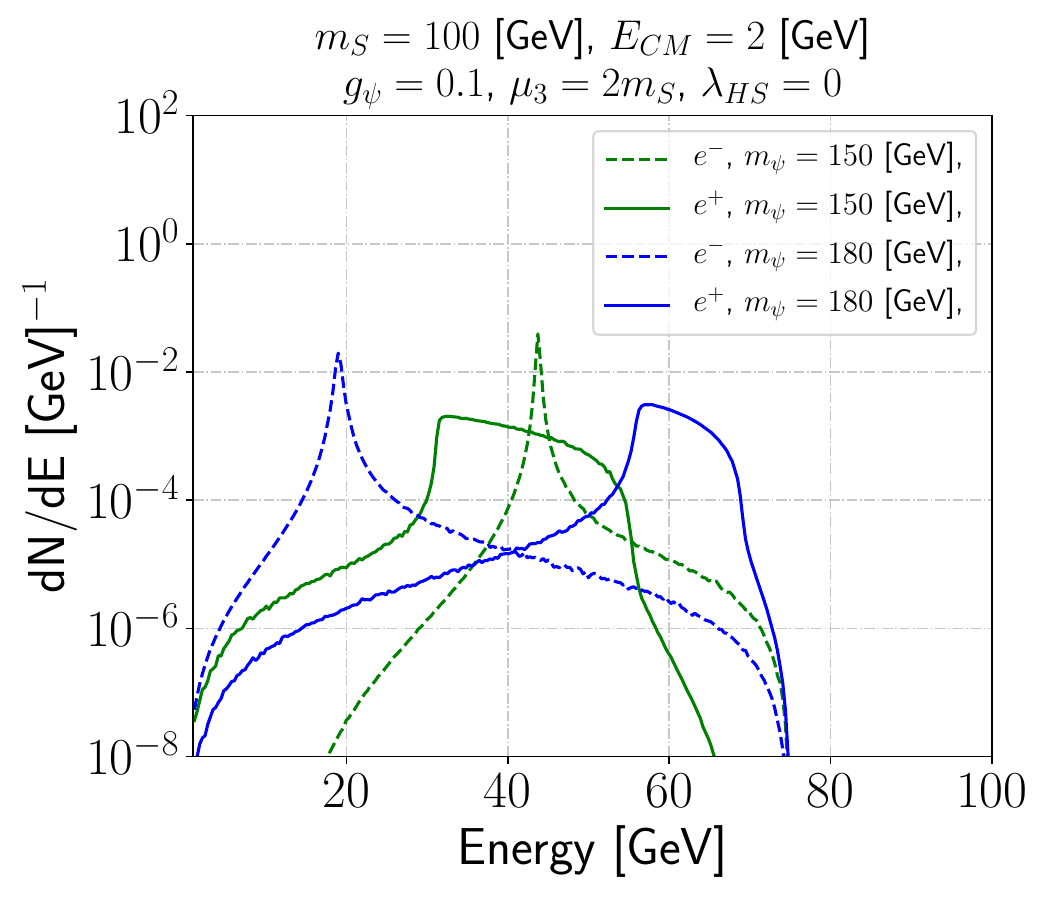}
\caption{\textit{Normalized spectra for the process $S + S \rightarrow e^+ + e^- + S^*$ as a function of the electron/positron energy, with the parameters indicated in the figure. The center-of-mass energy is given by $E_{\text{CM}}$.}} 
\label{spectra}
\end{figure}

In this appendix we detail some of the kinematics for $e^\pm$ produced from the complex scalar DM annihilation. In order to exemplify the typical kinematics occurring in this model in the mass range $m_\psi/2 < m_S < m_\psi$, we show in Fig.~\ref{spectra} the normalized spectra $\frac{dN}{dE} = \frac{1}{\sigma}\frac{d\sigma}{dE}$ for the process $SS\rightarrow e^-e^+S^*$ obtained with CalcHEP code \cite{Belyaev:2012qa}, assuming a center-of-mass momentum of $2$ GeV, and the parameters indicated in the plot. The bumps and sharp peaks can be understood by kinematics. For instance, the process $SS\rightarrow e^-\psi$ (all the argumentation afterwards goes also for the CP-conjugate processes), neglecting the electron mass, present the following energies for the final states 
\begin{eqnarray}\label{kine1}
 E_{e^-} = m_S \left(1 - \frac{m_\psi^2}{4m_S^2}\right), \quad E_{\psi} = m_S \left(1 + \frac{m_\psi^2}{4m_S^2}\right).
\end{eqnarray}
Considering $m_S = 100$ GeV and $m_\psi = 150$ GeV, we obtain in this $2\rightarrow 2$ process that $E_{e^-} \cong 43$ GeV (green sharp peak in Fig.~\ref{spectra}). Now, the energy distribution of $e^-/e^+$ is much more complex than the energies given in \ref{kine1}, as it can be seen in Fig.~\ref{spectra}, since this continuous spectra for the outgoing particles comes from the fact that the complete process is $2\rightarrow 3$ (see diagrams in Fig.~\ref{diagrams2}), with the diagrams interfering. However, other interesting information can be obtained considering the following. If we consider only the $s$-channel diagram, and assuming that the emitted $e^+$ from the $SS$ annihilation process makes a $\theta$ angle with respect to the in-flight $\psi$, the energy of the positron is given by
\begin{eqnarray}\label{kine0}
 E_{e^+} = \frac{(-m_S^2 + m_\psi^2)/2}{E_\psi - \sqrt{E_\psi^2 - m_\psi^2}\cos\theta},
\end{eqnarray}
with $E_\psi$ given in eq.~\ref{kine1}. Then, for $m_\psi = 150$ GeV, $E_{e^+} \approx [31, 56]$ GeV, which is exactly the place where the box-shaped spectra (continuous green line) is in the Fig.~\ref{spectra}. The maximum energy of $e^-/e^+$ is independent of $m_\psi$, and is defined uniquely by $m_S$. Equivalently, the previous analysis can be made for the case of $m_\psi = 180$ GeV (blue lines), in which certainly the energy distribution of $e^-/e^+$ changes at some extent, moving apart the electron peak from the box-shaped spectra of the positron.


\bibliography{bibliography}
\bibliographystyle{utphys}

\end{document}